\pdfoutput=1 

\documentclass[aps,pre,reprint,groupedaddress,twocolumn]{revtex4-1}

\usepackage[english]{babel}
\usepackage{amsmath,amssymb}
\usepackage{graphicx,color}
\usepackage{mathrsfs}
\usepackage{paralist}

\newcommand{\erf}{\operatorname{erf}}
\newcommand{\post}{\mbox{\tiny{post}}}

\newcommand{\vek}[1]{#1}
\newcommand{\pa}{\partial}
\newcommand{\lan}{\langle}
\newcommand{\ran}{\rangle} 
\newcommand{\lla}{\left\langle} 
\newcommand{\rra}{\right\rangle}

\newcommand{\pr}{p_p(x)}
\newcommand{\lik}{p_l(x)}
\newcommand{\llik}{\ln\lik}
\newcommand{\blik}{p_l^{\beta}(x)}
\newcommand{\dd}{\mbox{d}}
\newcommand{\dfn}{\mathrel{\mathop:}=}
\newcommand{\ppv}{p(d\vert \mathfrak{M})}

\begin{document}


\title{Prior-predictive value from fast-growth simulations: Error
analysis and bias estimation}



\author{Alberto Favaro}
\email[]{alberto.favaro@uni-oldenburg.de}
\author{Daniel Nickelsen}
\author{Elena Barykina}
\author{Andreas Engel}
\affiliation{Institut f\"ur Physik, Carl-von-Ossietzky-Universtit\"at, 26111 Oldenburg, Germany}

\date{\today}

\begin{abstract}
Variants of fluctuation theorems recently discovered in the statistical mechanics of non-equilibrium processes may be used for the efficient determination of high-dimensional integrals as typically occurring in Bayesian data analysis. In particular for multimodal distributions, Monte-Carlo procedures not relying on perfect equilibration are advantageous. We provide a comprehensive statistical error analysis for the determination of the prior-predictive value in a Bayes problem building on a variant of the Jarzynski equation. Special care is devoted to the characterization of the bias intrinsic to the method. We also discuss the determination of averages over multimodal posterior distributions with the help of a variant of the Crooks theorem. All our findings are verified by extensive numerical simulations of two model systems with bimodal likelihoods. 
\end{abstract}


\maketitle


\section{Introduction}
Statistical data analysis is at the heart of all quantitative science.
Observations, measurements, and numerical simulations alike are prone to
random perturbations, and effort and care is needed to scrutinize the
influence of these noisy disturbances on the results of the respective
investigation. A particularly clear and efficient procedure to do so is
provided by Bayesian inference
\cite{gelman1995,leonhard1999,jaynes2003}. In a typical setup, a model 
$\mathfrak{M}$ specified by parameters $x$ is checked against
observational, experimental or numerical data $d$. All information on
the parameters already available from previous experience is subsumed in
the {\em prior} distribution $p_{p}(x\vert \mathfrak{M})$ of the
parameters. The model itself is characterized by a {\em likelihood} distribution
$p_{l}(d\vert x,\mathfrak{M})$ specifying the probability of data
conditioned on a particular choice of the parameters. The application of
Bayes rule,
\begin{equation}\label{eq:bayes}
p_{\post}(x\vert d,\mathfrak{M})=\frac{p_{p}(x\vert
\mathfrak{M})p_{l}(d\vert x,\mathfrak{M})}{p(d\vert \mathfrak{M})} \,,
\end{equation}
then yields the {\em posterior} distribution $p_{\post}(x\vert
d,\mathfrak{M})$ for the parameters $x$. It provides the statistically
optimal combination of the information about the parameters contained in
the prior and in the new data. Bayesian methods are being used for
various problems in quite diverse fields of research
\cite{dose2003,dagostini2003,bernardo2011,vontoussaint2011}. They are in
particular appropriate for testing null-hypotheses \cite{anderson1992}
and in problems of model selection \cite{leonhard1999}. 

A crucial problem in concrete applications of Bayesian inference is the
determination of the denominator in \eqref{eq:bayes}, the so-called {\em
evidence} or {\em prior-predictive value}
\begin{equation}\label{eq:defppv}
p(d\vert \mathfrak{M})\dfn\int \hspace{-1.1pt}p_{p}(x\vert
\mathfrak{M})p_{l}(d\vert x,\mathfrak{M})\hspace{1pt}\mbox{d}^{n}x \,.
\end{equation}
Typically, the integral extends over a high-dimensional parameter space,
and is dominated by contributions from small and labyrinthine regions.
This makes straight Monte-Carlo methods rather inefficient
\cite{vonderlinden1999}. Since similar problems arise in statistical
mechanics in connection with the numerical determination of partition
functions or, equivalently, free energies, it is not surprising that
methods developed in statistical physics are being increasingly used in
data analysis. A prominent example is thermodynamic integration
\cite{kirkwood1935} which is meanwhile routinely implemented in Bayesian
inference \cite{neal1993,vonderlinden1999,vontoussaint2011}. Its
applicability rests on the accurate determination of thermal averages of
the logarithm of the likelihood distribution. This is a standard problem in
computational physics and can often be accomplished by Markov chain
Monte Carlo methods \cite{newman2006,vontoussaint2011}. Nevertheless, for
{\em multimodal} distributions, the relaxation times to thermal
equilibrium can be very large which may compromise the determination of
the necessary averages. In fact, for a model system with a bimodal
likelihood distribution, thermodynamic integration was shown to have substantial 
difficulties in determining the prior-predictive value of a Bayes problem \cite{ahlers2008}.

There are several situations in which multimodal distributions occur
quite naturally in Bayesian inference. A well-documented case is the
determination of the relative phase between two interferometers in the
presence of noise \cite{Stockton}. Plotting the two sinusoidal signals
against each other results in an ellipse, the ellipticity of which
determines the relative phase. Given the additional constraints present,
there remain {\em two} possible ellipses for each data point; the
corresponding likelihood distribution is hence bimodal. More complex 
situations are {\em mixture} models which allow for an
arbitrary number of components \cite{MaRo}. Problems of Monte-Carlo
methods for such mixture models are discussed, e.g., in \cite{Chopin2012}.

In recent years, there have been fascinating developments in the
statistical mechanics of non-equilibrium systems that gave rise to the
emerging field of stochastic thermodynamics
\cite{jarzynski2011,seifert2012,esposito2012}. Central to this field are
so-called work and fluctuation theorems which, among other things, may
be used to determine free-energy differences from non-equilibrium
trajectories \cite{liphardt2002,collin2005,pohorille2010}. Because of
the close relation between free-energy estimates and the calculation of
the prior-predictive value, these developments also bring about new
possibilities for Bayesian data analysis \cite{ahlers2008}. In an
inference problem, the non-equilibrium aspect is exhibited by the use of
{\em non-stationary}, explicitly time-dependent Markov processes which do
not rely on repeated equilibrations. Accordingly, when multimodal distributions
are considered, these methods can prove advantageous.

In the present paper, we analyze in detail the performance of an
algorithm to determine the prior-predictive value using a variant of the
Jarzynski equation \cite{jarzynski1997,jarzynski2004} that was proposed
in \cite{ahlers2008}. Of central importance in this connection is a
reliable error estimate of the method. Due to the non-linearities
involved, the method has a bias which needs to be treated with care \cite{gore2003, zuckerman2003, ytreberg2004}. We also detail the calculation of averages over multimodal posteriors using a variant of the Crooks relation.  

The paper is organized as follows. In section \ref{sec:be}, we provide
the basic equations and fix the notation. In section \ref{sec:ppv}, we
present a detailed error analysis of the method for determining the
prior-predictive value. Section \ref{sec:ex} demonstrates the performance of the proposed error analysis by means of two examples; a bimodal likelihood distribution composed of two Gaussians \cite{vonderlinden1999}, and a similar likelihood distribution but composed of two Lorentzians \cite{sivia1996}. Section
\ref{sec:post} provides an analogous analysis for averages with the
posterior distribution. Finally, section \ref{sec:conc} contains our
conclusions.


\section{Basic equations}\label{sec:be}
In the following, the dependency of the prior and the likelihood distribution on the parameters $x$ of the model is the important one. We therefore temporarily suppress the dependence on $d$ and $\mathfrak{M}$ for notational convenience.

For a successful application of Bayesian inference in problems of practical relevance, effective numerical methods are crucial. It is well-known that normalization factors of distributions like the prior-predictive value (PPV) are much harder to get by Monte-Carlo methods than the corresponding averages \cite{newman2006}. It were therefore desirable to replace the integration in \eqref{eq:defppv} by functions of such averages. A simple method to do so is the following variant of {\it thermodynamic integration} \cite{kirkwood1935}.  

Defining the auxiliary quantity
\begin{equation}\label{eq:defZ}
Z(\beta)\dfn\int \big(\lik\big)^\beta \pr \hspace{1pt}\dd^{n}x\, ,
\end{equation}
we have $Z(0)=1$ due to the normalization of the prior distribution and $Z(1)=\ppv$ which is the desired PPV. Moreover, 
\begin{equation}\label{eq:hti}
  \frac{\dd}{\dd \beta}\ln Z(\beta)=\frac{1}{Z(\beta)}\int \llik
  \;\blik \,\pr\hspace{1pt}\dd^{n}x\, .
\end{equation}
The r.h.s. of this equation denotes the average $\lan\llik\ran_\beta$ of the log-likelihood distribution with
\begin{equation}\label{eq:defPbeta}
   P_\beta(x)\dfn\frac{1}{Z(\beta)}\; \blik\,\pr \, . 
\end{equation} 
Hence,
\begin{equation}\label{eq:thdint}
  \begin{split}
    \ln\ppv&=\int_0^1 \dd\beta\, \frac{\dd}{ \dd\beta}\ln Z(\beta) \\
        &=\int_0^1 \dd\beta\, \lan \llik\ran_\beta \, .
  \end{split}
\end{equation}
In practical applications of this relation, one chooses $n=10...100$ values $\beta_n$ from the interval $(0,1)$ and calculates the averages $\lan\llik\ran_{\beta_n}$ by standard Markov chain Monte-Carlo (MCMC) sampling. The implemented transition probability $\rho(x,x';\beta_n)$ of the Markov chain has to be consistent with the corresponding stationary distribution \eqref{eq:defPbeta}. This is most directly ensured by the detailed balance condition \cite{newman2006}. Having obtained the $n$ averages $\lan\llik\ran_{\beta_n}$, the integral in \eqref{eq:thdint} can be determined approximately. We note that the Markov chain used for each of the $\beta$-values is stationary, i.e. there is no explicit time dependence in the transition probability $\rho(x,x';\beta_n)$.

This variant of thermodynamic integration works fine as long as there are no difficulties with the equilibration of the individual Monte-Carlo runs \cite{vonderlinden1999}. However, for multimodal distributions, problems may arise due to trajectories getting stuck in local maxima of the distribution \cite{Chopin2012}. In the generic case of unimodal prior and multimodal likelihood distributions, such problems show up when $\beta$ approaches 1. The last points for the calculation of the integral in \eqref{eq:thdint} are then prone to errors, and the whole estimate for the PPV becomes unreliable.  

These equilibration problems may be circumvented by building on modern methods for free-energy estimation that use non-stationary trajectories \cite{jarzynski1997,pohorille2010}. To this end, one considers a finite time interval $t\in(0,T)$ in which $\beta$ changes from 0 to 1. In the numerics, this is done by fixing a set of intermediate times and corresponding increments $\{t_m,\Delta\beta_m\}$, the so-called protocol $\beta(t)$. Starting from a point $x_0$ sampled from the prior distribution, MCMC simulations with the time-dependent transition rate $\rho(x,x';\beta(t))$ are performed. For each realization $x(t)$ of such a simulation, one determines the quantity
\begin{equation}\label{eq:defR}
 R[x(\cdot)]=\sum_m \Delta \beta_m \,\ln p_l(x(t_m))\,.
\end{equation}  
As shown in \cite{ahlers2008}, one then finds
\begin{equation}\label{eq:Jest}
 \lan e^R \ran=Z(1)=\ppv\, ,
\end{equation} 
where the average in \eqref{eq:Jest} is over independent realizations $x(t)$ of the non-stationary Markov process. In non-equilibrium thermodynamics, the above relation is known as the {\it Jarzynski equation}.\\
The continuum version of \eqref{eq:defR} has the form (see also \cite{Chernyak2007,Chatelain2007,Williams2008})
\begin{equation}\label{eq:defR_int}
 R[x(\cdot)]=\int_0^T\!\!\frac{\pa}{\pa t}\, \beta(t) \,\ln p_l(x(t))\, \dd t\,.
\end{equation}

Commonly, instead of $\ppv$, the logarithm of $\ppv$ is considered. This is due to several reasons. First, $\ppv$ is typically a very small number, entailing range errors in numerical operations. Second, the result of thermodynamic integration is already the log-PPV, see (\ref{eq:thdint}). And third, since the Jarzynski equation (\ref{eq:Jest}) is prominently used to estimate free-energies $\mathcal{F}=\ln Z$, existing results on error analysis in the determination of $\mathcal{F}$ can be adapted to the estimation of $\ln\ppv$ using the Jarzynski equation. Therefore, in this paper, we also will address $\ln\ppv$ instead of $\ppv$.

Remarkably, averages with the posterior distribution may be expressed in a similar way. For a reasonable function $f(x)$, one can show that \cite{Crooks2000,ahlers2008}
\begin{equation}\label{eq:postav}
 \langle f\rangle_{\mathrm{post}}=
   \int f(x)\,p_{\post}(x)\,  \dd x
   =\frac{\langle e^R f(x(T))\rangle}{\langle e^R \rangle}\, ,
\end{equation} 
where $x(T)$ denotes the final point of the trajectory $x(t)$, and the averages are again over an ensemble of realizations. 

\section{Error analysis of the Jarzynski estimator}\label{sec:ppv}
The Jarzynski equation (\ref{eq:Jest}) to determine the PPV $\ppv$ from non-stationary realizations $x(t)$ involves the exponential average 
\begin{equation} \label{eq:def_avg_exact}
  \lan e^R \ran \dfn \int \dd x\, p(R)\, e^R \,.
\end{equation} 
In practice, the distribution $p(R)$ of the random variable $R$ is unknown, and $\lan\dots\ran$ is replaced by an ensemble average,
\begin{equation} \label{eq:def_avg_sample}
  \lan e^R \ran_M \dfn \frac{1}{M}\sum\limits_{i=1}^{M} e^{R_i} \,,
\end{equation}
where the index $M$ in $\lan e^R \ran_M$ denotes the number of samples $R_i$ that contribute to $\lan e^R \ran_M$.\\
Replacing the exact average (\ref{eq:def_avg_exact}) with the sample mean (\ref{eq:def_avg_sample}) introduces an error to the Jarzynski estimator which vanishes in the limit of infinitely many samples, $M\to\infty$. However, due to the exponential weight on large $R$ values invoked by the non-linear average, this error may remain significant even for large $M$ \cite{gore2003,zuckerman2002,pohorille2010}. The analysis of this error is the central subject of this paper and will be discussed in this section.
 

\subsection{Basic notions}
To compute the log-PPV from a $M$-sized ensemble of $R$-values, we use (\ref{eq:Jest}) and (\ref{eq:def_avg_sample}) to define the {\it Jarzynski estimator}
\begin{equation} \label{eq:Jest_ln}
  \ln \ppv \simeq \ln \lan e^R \ran_M = \ln \frac{1}{M}\sum\limits_{i=1}^{M} e^{R_i} \,.
\end{equation}
Considering several $M$-sized ensembles of $R$-values, the sample mean $\lan e^R \ran_M$ is a random variable for any finite $M$. The statistics of $\ln \lan e^R \ran_M$ is central to our error analysis of the Jarzynski estimator. To assess the statistics of $\ln \lan e^R \ran_M$, we define bias $B$, variance $\sigma^2$ and mean square error $\alpha^{2}$ as
\begin{align}
B(M)&\dfn\lla\,\ln\lan e^{R}\ran_M\,\rra-\ln \ppv\,,\label{eq:bias}\\
\sigma^{2}(M)&\dfn\lla\,\bigl(\,\ln \langle e^{R}\rangle_M-\lla\,\ln \langle e^{R}\rangle_M\,\rra\,\bigr)^2\,\rra\,,\label{eq:var}\\
\alpha^{2}(M)&\dfn\lla\,\bigl(\ln \langle e^{R}\rangle_M-\ln \ppv\bigr)^{2}\,\rra\,. \label{eq:mse}
\end{align}
It is worth noting that these quantities are related by 
\begin{equation}
\alpha^{2}(M)=\sigma^{2}(M)+B^{2}(M)\,.\label{eq:biastomse}
\end{equation}

To understand why a non-zero bias \eqref{eq:bias} may occur, a valid starting point is
\begin{equation}
\lla\,\langle e^{R}\rangle_M\,\rra=\lla\, e^{R}\,\rra = \ppv\,.
\end{equation}
One substitutes this identity into the definition for the bias \eqref{eq:bias}, and establishes that
\begin{equation}
B(M)=\lla\,\ln \langle e^{R}\rangle_M\,\rra-\ln \lla\,\langle e^{R}\rangle_M\,\rra\,. \label{eq:finitebias}
\end{equation}
Hence, a finite bias signals that the logarithm and the expectation value do not commute. 

A related statement can be derived from the Jensen inequality \cite{billingsley1995}. If the function $\varphi$ is convex on the interval $I$, and $X$ is a stochastic variable with range $J \subseteq I$, then
\begin{equation}
\lla\,\varphi(X)\,\rra\geq\varphi(\lla X\rra)\,. \label{eq:jensen}
\end{equation}
When $\varphi(X)=-\ln(X)$ and $X=\langle e^{R}\rangle_M$, the inequality \eqref{eq:jensen} prescribes that 
\begin{equation}
\lla\,\ln\langle e^{R}\rangle_M\,\rra\leq \ln \lla\,\langle e^{R}\rangle_M\,\rra\,. 
\end{equation}
Thus, according to \eqref{eq:finitebias}, the bias of the Jarzynski estimator is negative, or zero. For the analogous property in statistical physics, we refer to \cite{jarzynski1997a, zuckerman2002}. 


\subsection{Confidence interval}\label{sec:ci}
If the bias $B$ and the variance $\sigma^2$ as defined in (\ref{eq:bias}) and (\ref{eq:var}) are known, the root mean square error $\alpha$ follows from (\ref{eq:biastomse}) and serves as a measure of uncertainty for the estimation of the log-PPV, $\ln\ppv$. While the computation of $\sigma^2$ from finite samples is straightforward, the determination of $B$ is intricate as it involves $\ppv$ itself. It therefore is common practice to substitute $\ppv$ with an appropriate estimator, in the case at hand being $\ppv\simeq\lan e^R \ran_M$. The consequence is that the resulting $\alpha$ only accounts for the bias generated by the logarithm in the Jarzynski estimator (\ref{eq:Jest_ln}), and not for the non-linearity of the exponential average. In what follows, the full bias $B$ will be split into two contributions $C$ and $D$, in which $C$ uses the mentioned substitution $\ppv\simeq\lan e^R \ran_M$, and $D$ takes care of the error brought about by this step.

In tackling the intrinsic problem that the true values of $\ppv$, and therefore also $B$, are not known, the key point will be to derive a confidence interval for $D$ from the central limit theorem \cite{billingsley1995}. To do so, we make two assumptions:
\begin{enumerate}[(i)]
\item $\{e^{R_{1}},\dots,e^{R_{\hspace{-0.5pt}N}}\hspace{-1pt}\}$ is a sequence of $N$ independent random variables that have the same distribution;\label{en:iid}
\item the variance $\varsigma^{2}$ of that distribution is finite -- while the expectation value is $\ppv$, because of \eqref{eq:Jest}. \label{en:var}
\end{enumerate}
The sample size $N$, in addition to $M$, is introduced for later convenience, and we assume that $N\gg M$.
Note that (\ref{en:var}) refers to the distribution of $e^{R}$, the variance of which may be finite despite likelihood distributions with infinite variance. We will demonstrate this point in section \ref{sec:doublecauchy}.

If (\ref{en:iid}) and (\ref{en:var}) are satisfied, the central limit theorem dictates that, as $N$ approaches infinity, the random variable
\begin{equation}
Y(N)\dfn\sqrt{N}\bigl(\langle e^{R}\rangle_N-p(d\vert\mathfrak{M})\bigr) {\big/} {\varsigma} \label{eq:appzscore}
\end{equation}
becomes normally distributed, with zero mean and unit variance. Accordingly, a confidence interval for $Y(N)$ may be written as
\begin{equation}
\Pr\Bigl[-\sqrt{2}\erf^{-1}\hspace{-1pt}(\gamma)<Y(N)<\sqrt{2}\erf^{-1}\hspace{-1pt}(\gamma)\Bigr]\approx\gamma\,,\label{eq:ciy}
\end{equation}
where $\Pr[\dots]$ indicates probability, and $\erf^{-1}$ is the inverse error function. The confidence level $\gamma$ can be selected as one deems fit, but ordinary choices are $95\%$, $99\%$, $99.5\%$, and $99.9\%$, see \cite{harnett1982}. Throughout this paper, we will use the rather pessimistic choice $\gamma=0.95$. The approximate sign in \eqref{eq:ciy} accounts for the fact that $N$ is taken to be finite.

The confidence interval for $Y(N)$ can be transferred to the bias $B$. To this end, we solve \eqref{eq:appzscore} for $\ppv$, and substitute the result into \eqref{eq:bias}. Hence, the bias is expressed as
\begin{align}
B(M)&=C(M,N)+D(N)\,,\label{eq:cplusd}
\end{align}
with 
\begin{align}
C(M,N)&=\lla\ln \langle e^{R}\rangle_M\rra-\ln\langle e^{R}\rangle_N\,,\label{eq:c}\\
D(N)&=-\ln\bigg[1-\frac{{\varsigma}}{\sqrt{N}}\frac{Y(N)}{\langle e^{R}\rangle_N}\bigg]\,.\label{eq:d}
\end{align}
The dependency on $N$ of the first term in (\ref{eq:cplusd}) is compensated by the second term. We multiply the inequality within brackets in \eqref{eq:ciy} by the positive quantity ${\varsigma}/\sqrt{N}\langle e^{R}\rangle_N$. Then, to incorporate $D(N)$, we apply the monotonic increasing function $-\ln(1-X)$, see \eqref{eq:d}. Both these operations do not reverse the sign of the inequality. It follows that
\begin{equation}
\Pr\Bigl[D_{-}(N)<D(N)<D_{+}(N)\Bigr]\approx\gamma\,,\label{eq:cid}
\end{equation}
with 
\begin{equation}
D_{\pm}(N)=-\ln\bigg[1\mp\sqrt{\frac{2}{N}}\frac{{\varsigma} \erf^{-1}\hspace{-1pt}(\gamma)}{\langle e^{R}\rangle_N}\bigg]\,.\label{eq:dpm}
\end{equation}
Finally, by adding $C(M,N)$ to the inequality within brackets in (\ref{eq:cid}), a confidence interval for the bias is attained, 
\begin{equation} \label{eq:cib}
\Pr\Bigl[B_{-}(M,N)<B(N)<B_{+}(M,N)\Bigr]\approx\gamma\,,
\end{equation}
with the confidence limits 
\begin{equation}
B_{\pm}(M,N)=C(M,N)+D_{\pm}(N)\,.\label{eq:bpm}
\end{equation}

Two comments are in order. {First}, when $N$ is large enough, one has that
\begin{equation}
0<\sqrt{\frac{2}{N}}\frac{\varsigma}{\langle e^{R}\rangle_N}<1\,.
\end{equation}
The inequality on the left always holds true. Because $\gamma$ is positive, $\erf^{-1}\hspace{-1pt}(\gamma)$ ranges from $0$ to $1$, and
\begin{equation}
0<\sqrt{\frac{2}{N}}\frac{{\varsigma} \erf^{-1}\hspace{-1pt}(\gamma)}{\langle e^{R}\rangle_N}<1\,.
\end{equation}
This ensures that the confidence limit $D_{+}(N)$ is finite and real, cf. \eqref{eq:dpm}. {Second}, in \eqref{eq:bpm}, the dependency of $C(M,N)$ on $N$ is not compensated by that of $D_{\pm}(N)$. It follows that the confidence limits $B_{\pm}(M,N)$ are functions of $M$ and also $N$.  

We are now in the position to derive a confidence interval for the mean square error $\alpha^{2}(M)$, see (\ref{eq:biastomse}). Motivated by the procedure followed earlier on, it is natural to define
\begin{align}
\alpha_{\pm}^{2}(M,N)&=\sigma^{2}(M)+B_\pm^{2}(M,N)\,. \label{eq:hatalpha2}
\end{align}
Unfortunately, this is not a monotonic function, and the direction of previous inequalities gets mixed up. Nevertheless, it is still possible to conclude that
\begin{equation}
\Pr\Bigl[\alpha^{2}(M)<\max\bigl[\alpha^{2}_{+}(M,N),\alpha^{2}_{-}(M,N)\bigr]\Bigr]\gtrsim \gamma\,,\label{eq:cimse}
\end{equation}
where $\max[\dots]$ selects the larger of its two arguments.

The error analysis proposed above involves the exact averages $\lla\dots\rra$. For practical purposes, however, it is necessary to estimate the averages $\lla\dots\rra$ by empirical averages as defined in (\ref{eq:def_avg_sample}). To do so, we take $N$ as the given total number of $R$-values, group these into $N/M$ blocks of size $M$, and estimate
\begin{align} \label{eq:block-avg}
	\lla \lan\dots\ran_M \rra \simeq \big\lan \lan\dots\ran_M \big\ran_{\frac{N}{M}} \,.
\end{align}
This procedure, commonly referred to as block-averaging, was pioneered by Wood, M\"uhlbauer and Thompson \cite{wood1991}. We mention that an alternative to block-averaging is the bootstrap algorithm, as explored in the article \cite{ytreberg2004} by Ytreberg and Zuckerman.\\
In the remaining part of the paper, we will use the prescription (\ref{eq:block-avg}) to estimate $C(M,N)$ and $\sigma^2(M)$, defined in (\ref{eq:c}) and (\ref{eq:var}), from simulation results of an $N$-sized ensemble of $R$-values. To estimate $D(N)$ and $D_\pm(N)$, defined in (\ref{eq:d}) and (\ref{eq:dpm}), as well as the confidence interval for the bias in (\ref{eq:cib}) and the mean square error in (\ref{eq:cimse}), we approximate the variance $\varsigma^2$ of the distribution for $e^R$ with the sample variance
\begin{align} \label{eq:varsigma_hat}
	\hat\varsigma^2(N) \dfn \frac{1}{N-1}\sum_{i=1}^{N}\bigl(e^{R_{i}}-\langle e^{R}\rangle_N\bigr)^{2} \,.
\end{align}
Likewise, for $\sigma^2$, we take
\begin{align} \label{eq:sigma_hat}
	\hat\sigma^2(M,N) \dfn \frac{1}{N/M-1}\sum_{i=1}^{N/M}\bigl(\ln \langle e^{R}\rangle_M-\big\lan\ln \langle e^{R}\rangle_M\big\ran_{\frac{N}{M}}\bigr)^2 \,.
\end{align}
We will denote estimated quantities that use block-averages and sample variances instead of exact averages with a 'hat', for instance,
\begin{align}
	\hat B(M,N) &= \big\lan\ln \langle e^{R}\rangle_M\big\ran_{\frac{N}{M}}-\ln\ppv\,,\label{eq:b_hat}\\
	\hat C(M,N) &= \big\lan\ln \langle e^{R}\rangle_M\big\ran_{\frac{N}{M}}-\ln\langle e^{R}\rangle_N\,,\label{eq:c_hat}\\
	\hat D_{\pm}(M,N) &= -\ln\bigg[1\mp\sqrt{\frac{2}{N}}\frac{{\hat\varsigma(N)} \erf^{-1}\hspace{-1pt}(\gamma)}{\langle e^{R}\rangle_N}\bigg]\,,\label{eq:dpm_hat}\\
	\hat\alpha^2(M,N) &= \hat\sigma^2(M,N) + \hat B^2(M,N)\,, \label{eq:mse_hat}
\end{align}
in contrast to the exact expressions (\ref{eq:d}), (\ref{eq:c}), (\ref{eq:dpm}) and (\ref{eq:biastomse}). The confidence limits $\hat D_{\pm}(M,N)$, as opposed to $D(N)$, are independent of the unknown $\ppv$. Accordingly, the same holds true for
\begin{equation}
\hat\alpha_{\pm}^2(M,N) = \hat\sigma^2(M,N) + \bigl(\hat C(M,N)+\hat D_{\pm}(M,N)\bigr)^{2}\,.
\end{equation}


\section{Examples for PPV estimation}\label{sec:ex}

Section \ref{sec:ppv} was devoted to the bias $B$ of the Jarzynski estimator (\ref{eq:Jest_ln}). We split the bias into two components, $B=C+D$, where $C$ is treated by block-averaging and $D$ is the remaining unknown discrepancy of the estimator. Based on the central limit theorem, we derived the confidence limits $D_\pm$ for the unknown $D$.\\
In this section, we demonstrate the performance of the Jarzynski estimator and the proposed error analysis for two exactly solvable settings involving bimodal likelihood distributions.
We also relate our error analysis to those existing in the literature, which exemplifies that $C$ is useful to judge the applicability of the central limit theorem, indicating the minimum total number of $R$-values for which $D_\pm$ becomes reliable.


\subsection{Gaussian bimodal likelihood distribution}\label{sec:simbim}
To construct a bimodal $P_{\beta}(x)$, the simplest option appears to be that of setting the likelihood distribution $p_{l}(d\vert x,\mathfrak{M})$ to be the sum of two Gaussians \cite{vonderlinden1999}. Hence, we specify that 
\begin{equation}
p_{l}(d\vert x,\mathfrak{M})=q_1G\bigl(x,d,\sigma_{l}^{2}\bigr)+q_2G\bigl(x,-d,\sigma_{l}^2\bigl)\,,\label{eq:exlike}
\end{equation}
where $x$ and $d$ are vectors of dimension $n$, and $q_1$ and $q_2$ assign different weights to the Gaussians
\begin{equation}
G\bigl(x,\mu,\sigma^{2}\bigr)=(2\pi\sigma^{2})^{-n/2} \exp\left(-\frac{(\mu-x)^{2}}{2\sigma^{2}} \right)
\end{equation} 
with mean-vector $\mu$ and variance $\sigma^2$. Choosing values for $q_1$ and $q_2$ that differ substantially from each other makes the equilibration problem particularly pronounced: while the positions of the maxima become apparent rather quickly, sampling the maxima with the correct weights $q_1$ and $q_2$ is reliant on the very rare trajectories that cross the low-probability region between the maxima.\\
The benefit of the Gaussian model is that the PPV is known analytically -- if the prior-distribution is taken to be Gaussian. Notably, this choice for $p_{p}(x\vert \mathfrak{M})$ is widespread in the Bayesian inference literature. Thus, we demand that 
\begin{equation}
p_{p}(x\vert \mathfrak{M})=G\bigl(x,0,\sigma_{p}^{2}\bigr)\,. \label{eq:exprio}
\end{equation}
to find 
\begin{equation}
\ppv=G\bigl(d,0,\sigma^{2}_{p}+\sigma_{l}^{2}\bigr)\,. \label{eq:exappv}
\end{equation}
We therefore have an analytic result which we can use to test our error analysis.\\
The dimension $n$ will be set to $5$, and $d$ will be taken to have all of its components equal to $10$. The maxima of the likelihood distribution are hence located at $\pm d=\pm(d_1,\dots,d_5)=\pm(10,\dots,10)$. For the weights $q_i$ we choose $q_1=\frac{1}{21}$ and $q_2=\frac{20}{21}$. The variances in \eqref{eq:exlike} and \eqref{eq:exprio} are selected to be $\sigma_{l}^2=1$ and $\sigma_p^2=100$, since in the typical Bayesian setup, the prior distribution is much broader than the likelihood distribution.

As discussed in section \ref{sec:be}, the protocol $\beta$ varies from $0$ to $1$ along every trajectory. We prescribe that $\beta$ increases in a cubic way,
\begin{align} \label{eq:proto}
	\beta=0.05t + 0.95t^3 \,,
\end{align}
where $t$ is incremented from $0$ to $1$ in $25$ steps. For each value of the protocol, the MCMC algorithm explores the parameter space, with $20$ steps in the Markov chain. These values correspond to relatively short trajectories, whereby the computational resources can be focused in generating a large number $N$ of $R$-values.

In the article \cite{zuckerman2003}, Zuckerman and Woolf demonstrate that, when $M$ is large, 
\begin{equation}
-B(M)\approx \frac{\sigma^2(M)}{2}\approx \frac{1}{2M}\!\left[\frac{\varsigma}{\ppv}\right]^2, \label{eq:largem}
\end{equation}
as a consequence of the central limit theorem. The same result is obtained in the paper \cite{gore2003} by Gore, Ritort and Bustamante. It is worthwhile to observe that \eqref{eq:largem} involves only exact quantities. Accordingly, the above relation can be used to identify a threshold for $M$ above which the central limit theorem for the random variable $\lan e^R\ran_M$ may assumed to be applicable. As the derivation of the confidence limits $D_\pm(N)$ rests on this very assumption, we conclude that $D_\pm(N)$ becomes reliable for values of $N$ above the same threshold.

\begin{figure}
  \centering
    \includegraphics[width=0.95\columnwidth]{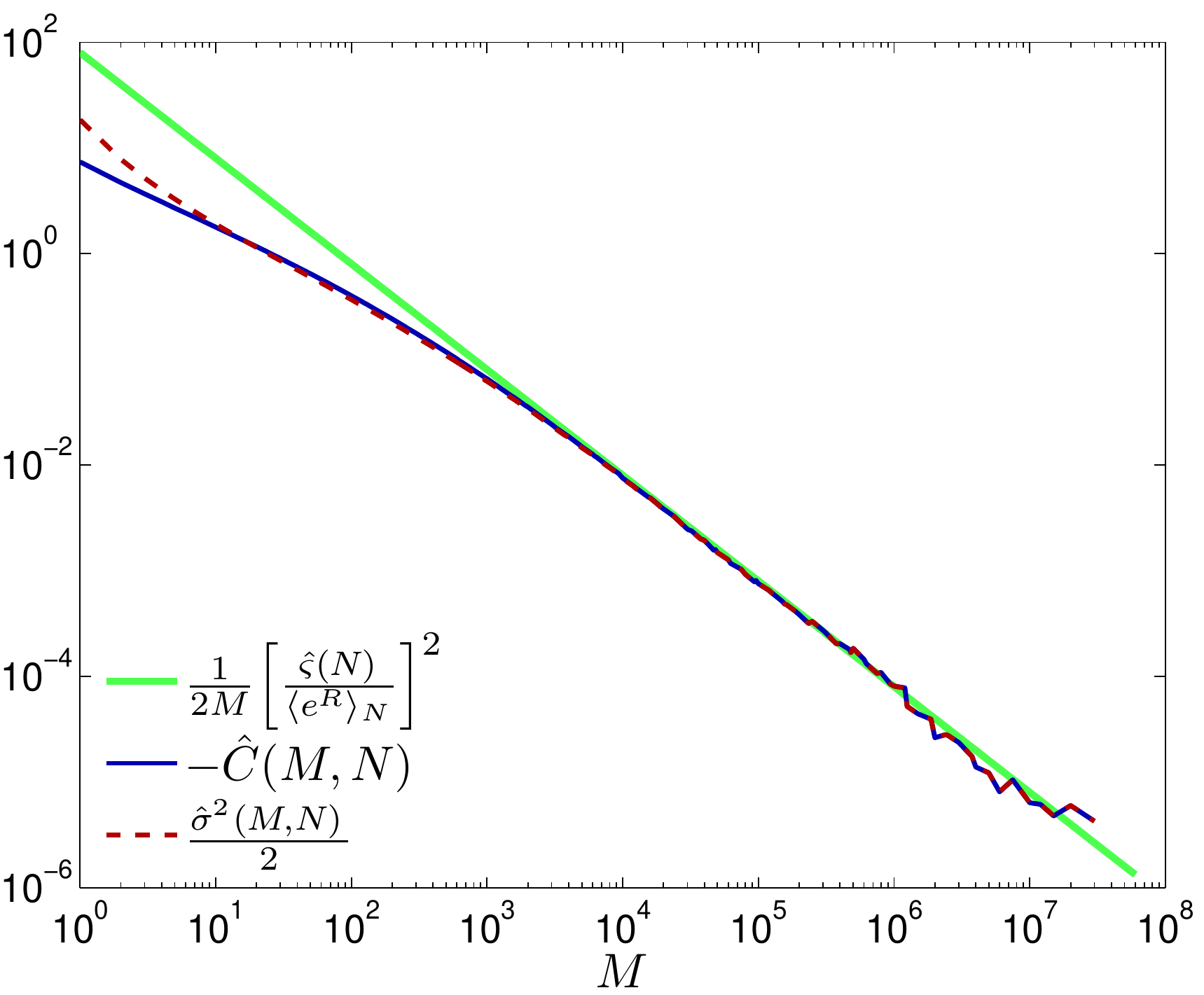}
    \caption{\label{fig:cmn} Verifying that our error analysis is consistent with the result \eqref{eq:largem_hat} obtained in \cite{zuckerman2003,gore2003} based on the central limit theorem for the random variable $e^R$. The total number of Markov chains, and consequently of $R$-values, is $N=6\times 10^{7}$.}
\end{figure}
To identify for the introduced bimodal Gaussian example (\ref{eq:exlike}) the regime where the central limit theorem is applicable, we generated a total number of $N=6\times10^7$ $R$-values and substitute these in the numerically accessible variant of (\ref{eq:largem}),
\begin{equation}
-\hat C(M)\approx \frac{\hat\sigma^2(M)}{2}\approx \frac{1}{2M}\!\left[\frac{\hat\varsigma(N)}{\lan e^R \ran_N}\right]^2\,, \label{eq:largem_hat}
\end{equation}
as used by Gore et al. in \cite{gore2003}. In Fig.\ \ref{fig:cmn} we plot the three quantities in (\ref{eq:largem_hat}) for all possible divisors $M$ of $N=6\times10^7$. The threshold above which the central limit theorem applies appears to be about $M\approx10^4$, for $M>10^4$ all quantities exhibit the predicted $1/M$ behavior. Therefore, our error analysis is in agreement with \cite{zuckerman2003,gore2003}.

\begin{figure}
  \centering
    \includegraphics[width=0.95\columnwidth]{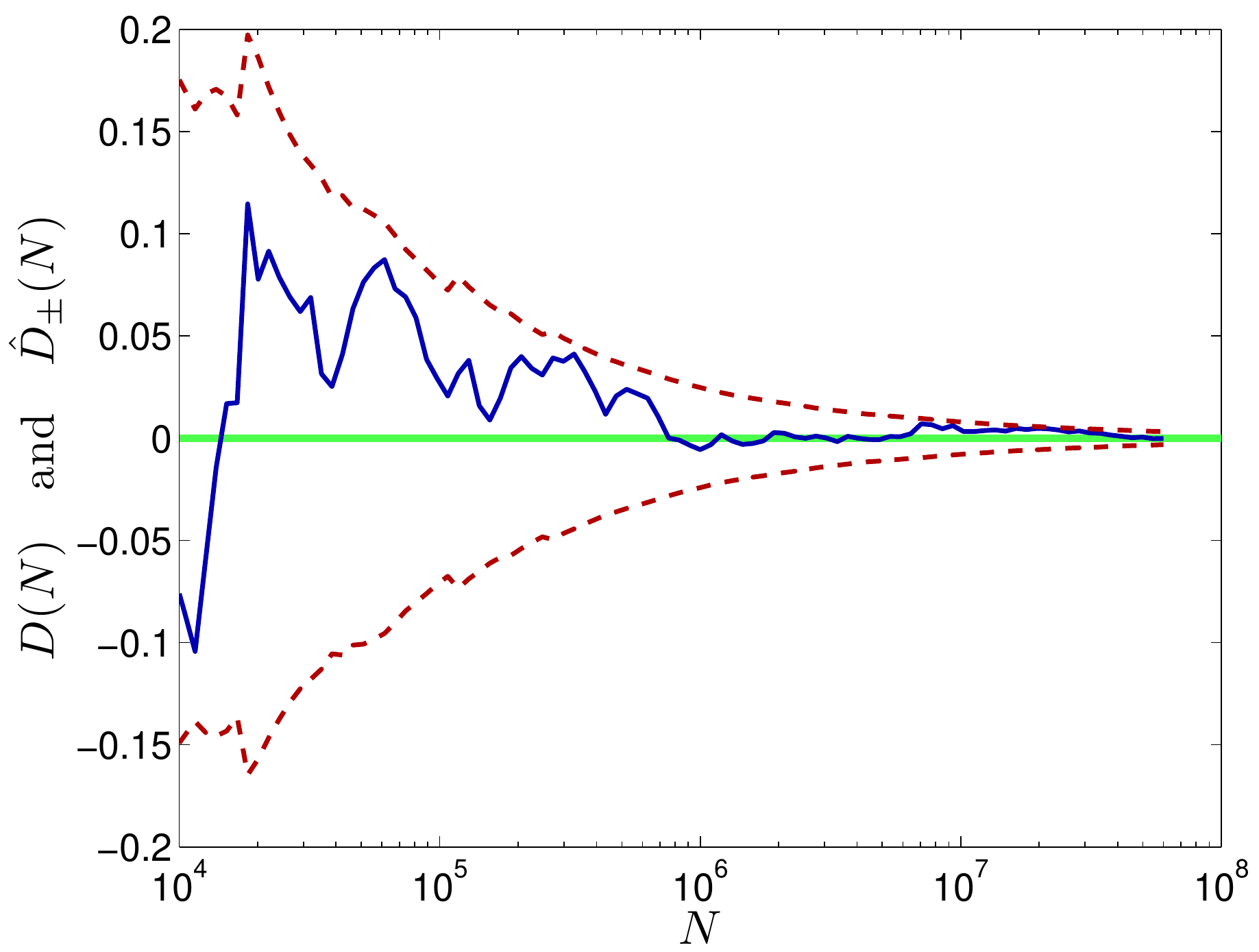}
    \caption{\label{fig:dn} As the number of $R$-values, $N$, gets larger, the confidence interval \eqref{eq:cid} gets smaller. Nevertheless, for the data set examined here, $D(N)$ stays within $\hat D_{+}(N)$ and $\hat D_{-}(N)$. The value of $D(N)$ follows from (\ref{eq:d}) using the exact value of $\ppv$ given by (\ref{eq:exappv}), the estimated confidence limits $\hat D_\pm(N)$ derive from (\ref{eq:dpm_hat}).}
\end{figure}
To gain an error margin for the estimation of $\ln\ppv$, one could choose $M<N$ and use $\lan\ln\lan e^R\ran_M\ran_{N/M}$ as an estimator, for which $C(M,N)$ is an appropriate error measure. However, the best estimate for $\ln\ppv$ is obtained by choosing $M=N$, i.e. $\ln\lan e^R\ran_N$, which is also signified by $C(N,N)=0$. The price we pay in using the best estimate is that the block-average procedure gives no statement for the uncertainty of the estimator. At this point, the remaining part of the bias, which we introduced as $D(N)$, enters the picture. In Fig.\ \ref{fig:dn}, we plot $D(N)$ by using the exact result for $\ppv$ in (\ref{eq:exappv}). In the usual case in which $\ppv$ is not known, one has to resort to the confidence limits $D_\pm(N)$, which we estimated by using (\ref{eq:dpm_hat}) and included into the plot. The depicted range of $N$-values is larger than the threshold $10^4$ above which we assume the central limit theorem for $\lan e^R \ran_N$ to hold and $D_\pm(N)$ to be reliable. Indeed, it is observed that $D_\pm(N)$ smoothly approach zero and that $D(N)$ belongs to the confidence interval \eqref{eq:cid}.

\begin{figure}
\includegraphics[width=0.95\columnwidth]{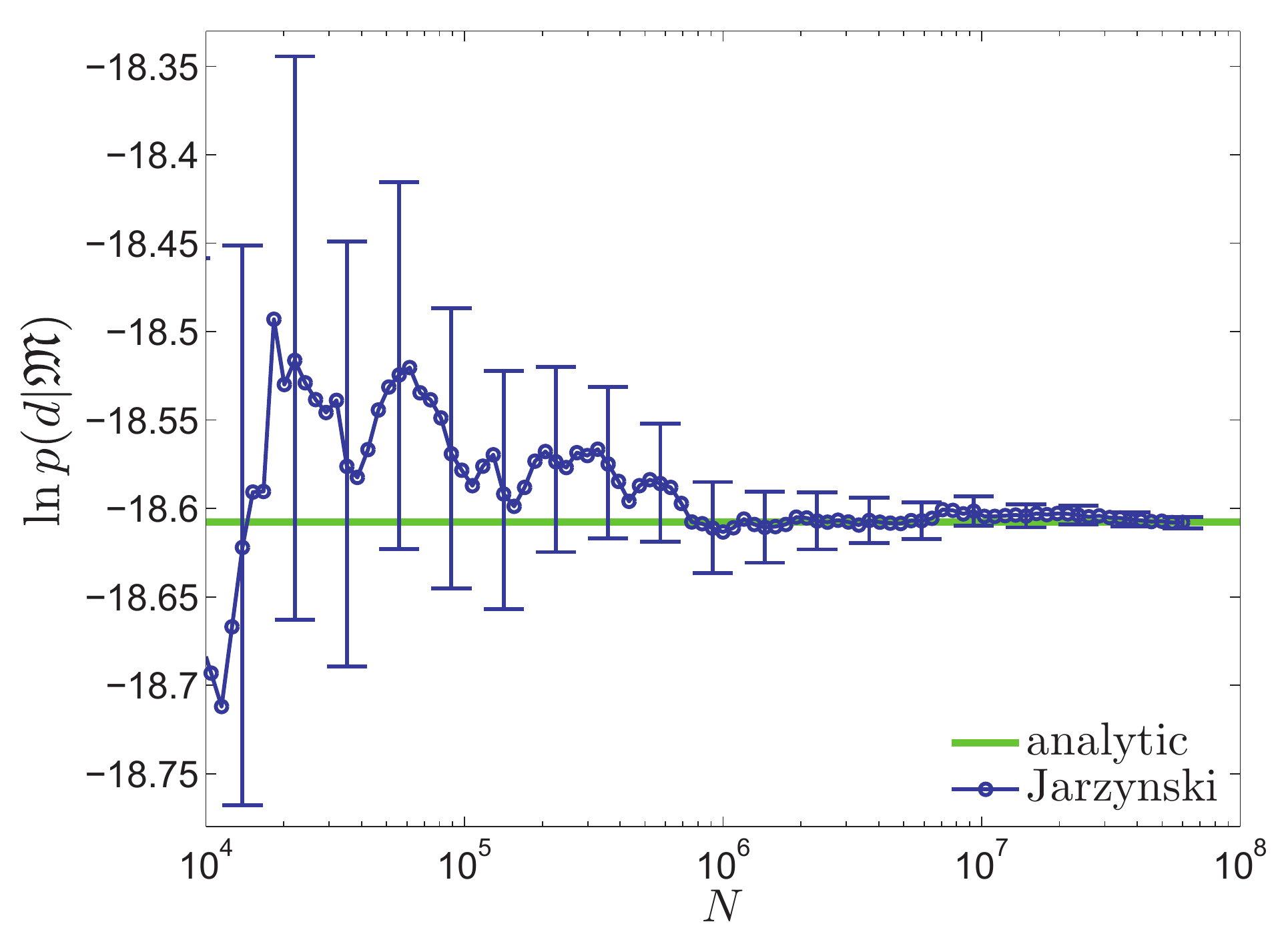}%
\caption{\label{fig:GAU_lnPPV_conv_Nrow1} Estimation of the log-PPV for the bimodal Gaussian model (\ref{eq:exlike}) using the Jarzynski estimator (\ref{eq:Jest_ln}) for an increasing number $N$ of $R$-values. The thick line is the analytic result (\ref{eq:exappv}), the symbols use the Jarzynski estimator. The error bars are given by the estimates of the bias, $\hat D_{\pm}(N)=\hat B_{\pm}(N,N)=\hat\alpha_{\pm}(N,N)$ from (\ref{eq:dpm_hat}).}
\end{figure}
Finally, in Fig.\ \ref{fig:GAU_lnPPV_conv_Nrow1}, we demonstrate the performance of the Jarzynski estimator and the proposed error analysis for an increasing number $N$ of considered $R$-values. For the best estimate, i.e. $M=N$ and $\hat\sigma(N)=0$, the estimated root mean square error is $\hat\alpha=\hat B$, and as furthermore $C(N,N)=0$, it is simply $\hat\alpha_{\pm}=\hat D_{\pm}(N)$. For the smallest value of $N$ we again choose the threshold $N=10^4$ above which the confidence limits $\hat D_\pm(N)$ are assumed to be reliable. We therefore use in Fig.\ \ref{fig:GAU_lnPPV_conv_Nrow1} the limits $\hat D_{\pm}(N)$ as error bars, which are found to always cover the analytic result.

\subsection{Likelihood distribution with infinite variance}\label{sec:doublecauchy}
The proposed error analysis in Sec.\ \ref{sec:ci} relies on the applicability of the central limit theorem to the random variable $\mathrm{e}^R$, that is, a finite variance $\varsigma^2$. As mentioned before, the requirement $\varsigma^2<\infty$ does not restrict to likelihood distributions with finite variance, which we demonstrate in this section.

To this end, we consider the Cauchy distribution (also known as a Lorentzian)
\begin{equation} \label{eq:FAT_cauchy}
  p(x) = \frac{s}{\pi}\Big[s^2 + (x-d)^2\Big]^{-1} \,.
\end{equation}
The moments of the Cauchy distribution do not exist, in particular the variance is divergent. Therefore, instead of mean and variance, the Cauchy distribution is characterized by the parameters $d$ and $s$, where $d$ is the mode of $p(x)$, and $s$ specifies the width, as $2p(d+s)=p(d)$.

The cumulative distribution is known analytically and reads
\begin{equation} \label{eq:FAT_cauchy_cum}
  P(x) = \frac{1}{\pi}\arctan\left[\frac{x-d}{s}\right] \,.
\end{equation}
To ensure a close analogy to the Gaussian example, we combine two Cauchy distributions to construct the bimodal likelihood distribution
\begin{align} \label{eq:FAT_cauchy_bimodal}
	\begin{split}
	p_{l}(d\vert x)= \left(\frac{s}{\pi}\right)^n\,\Bigg[ &q_1\prod_{i=1}^n \Big[s^2+(d_i-x_i)^2\Big]^{-1} \\ 
	&+ q_2\prod_{i=1}^n \Big[s^2+(d_i+x_i)^2\Big]^{-1} \Bigg] \,.
	\end{split}	
\end{align}
Here, $\vek{x}$ is a $n$-dimensional parameter-vector, and we take again one measurement $\vek{d}$ to be of the same dimension as $\vek{x}$.

Cauchy distributions are known to occur in power spectra of oscillating signals \cite{Sivia1992,sivia1996}. A recent example of a Bayesian analysis are helioseismic spectra to probe the interior of stars \cite{appourchaux1998,gruberbauer2009}, in which a multimodal likelihood distribution of the form (\ref{eq:FAT_cauchy_bimodal}) is used. For a limited number of data-points, the posterior is typically multimodal itself due to peaks in the power spectra being artifacts of data processing or of instrumental origin.

\begin{figure}
\includegraphics[width=0.95\columnwidth]{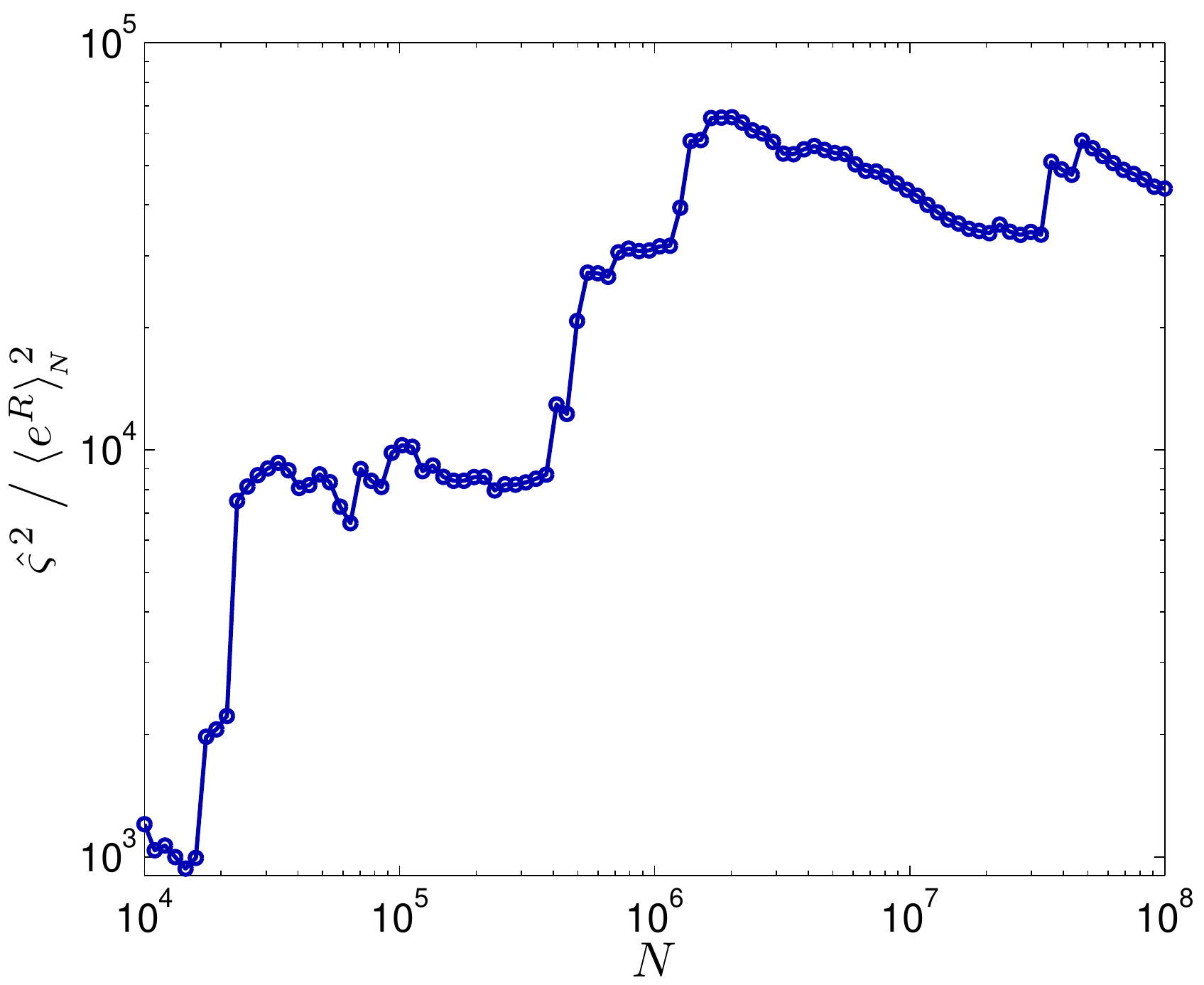}%
\caption{\label{fig:FAT_var_expR_conv} Estimated variance $\hat{\varsigma}^{2}(N)$ for $e^{R}$, when the likelihood distribution is the sum of two Cauchy distributions, see (\ref{eq:FAT_cauchy_bimodal}). We observe that $\varsigma^{2}$ remains finite, and thus the central limit theorem can be applied.}
\end{figure}
\begin{figure}
\includegraphics[width=0.95\columnwidth]{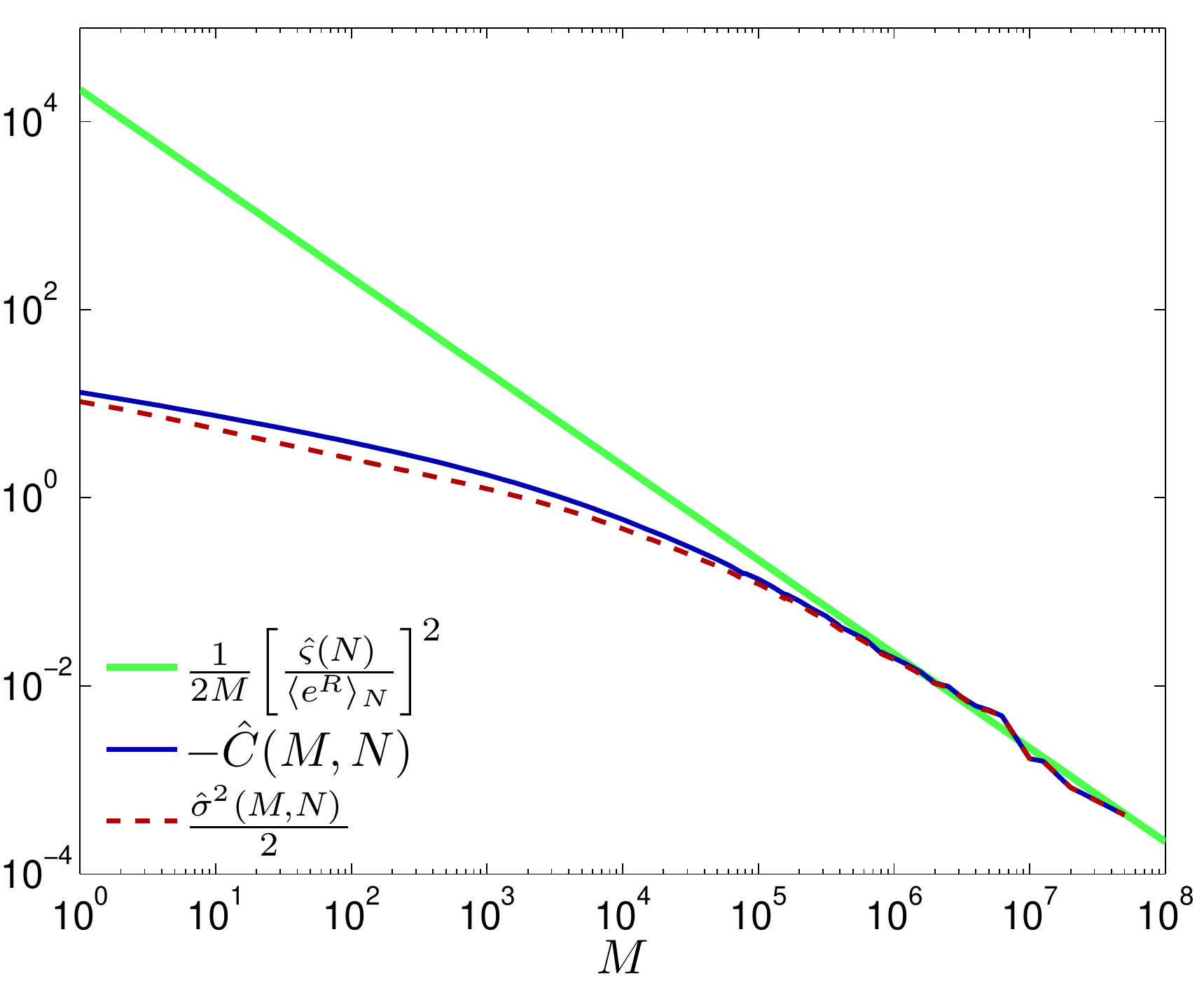}%
\caption{\label{fig:FAT_gore} Verifying for the likelihood distribution (\ref{eq:FAT_cauchy_bimodal}) involving two Cauchy distributions that our error analysis is consistent with the result \eqref{eq:largem_hat} obtained in \cite{zuckerman2003,gore2003} based on the central limit theorem for the random variable $e^R$. The total number of Markov chains, and consequently of $R$-values, is $N=10^{8}$.}
\end{figure}
Similar to the Gaussian example, we choose the parameters $n=5$, $d=(10,10,10,10,10)^\mathrm{T}$, $s=0.1$, $q_1=20/21$ and $q_2=1/21$. In order to compute the PPV analytically from (\ref{eq:FAT_cauchy_cum}), we employ a flat prior on the interval $[-20s, 20s]$. The interval covers both modes of the likelihood distribution and therefore does not include any a-priori information on the shape of the likelihood distribution; in fact, choosing a flat prior that does not cover the modes is found to drastically improve the performance of the Jarzynski method, since the Markov chains never start at one of the modes but instead run into the respective minima according to the weights $q_1$ and $q_2$. The protocol $\beta(t)$ is the same as for the Gaussian example, see (\ref{eq:proto}).

\begin{figure}
\includegraphics[width=0.95\columnwidth]{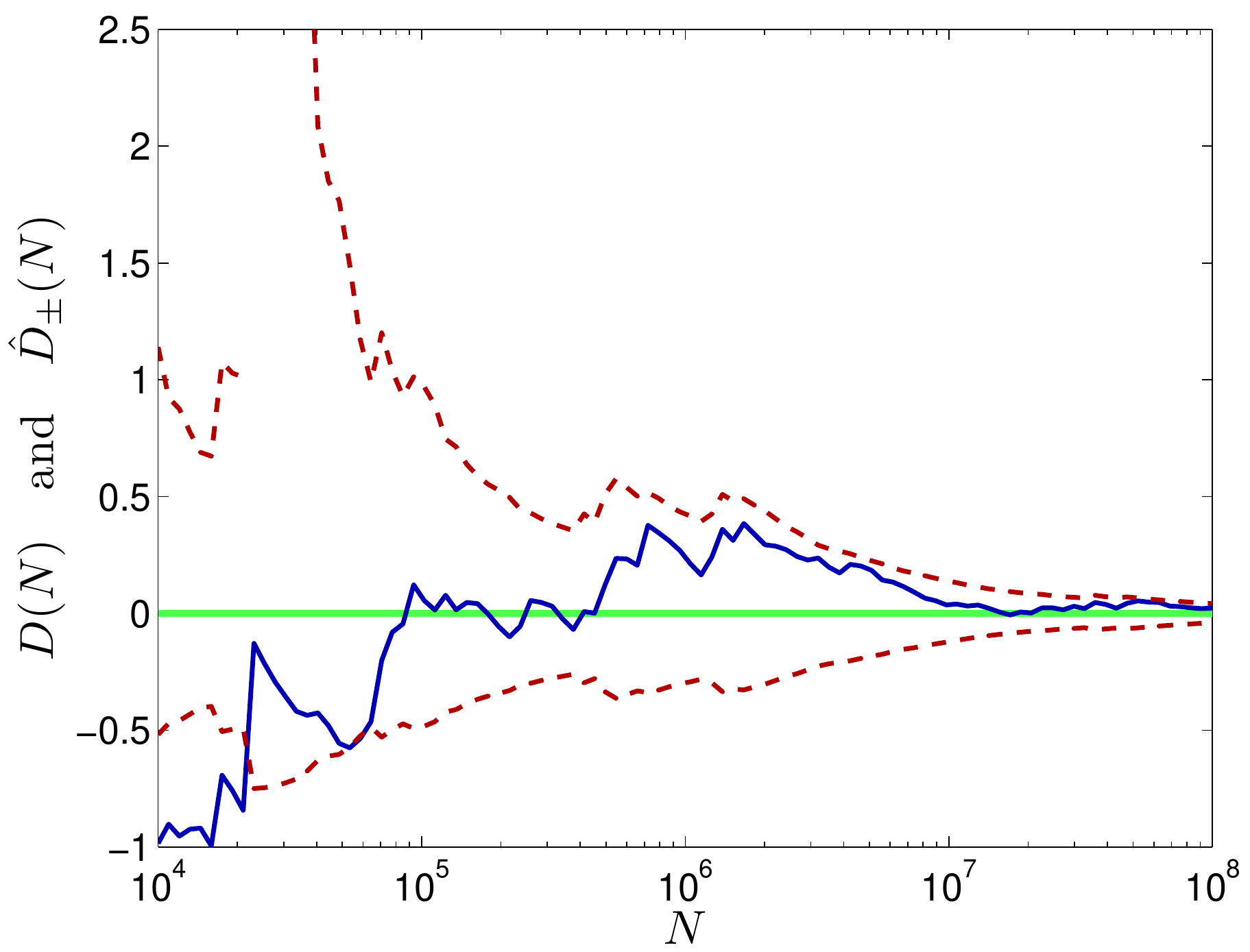}%
\caption{\label{fig:FAT_bias} The value of $D(N)$ defined in (\ref{eq:d}) together with its confidence interval given by $\hat D_\pm(N)$ from (\ref{eq:dpm_hat}) as a function of increasing number $N$ of considered samples for $R$ for the likelihood distribution in (\ref{eq:FAT_cauchy_bimodal}). The determination of $D(N)$ involves the exact value of $\ppv$ which follows from using the cumulative Cauchy distribution in (\ref{eq:FAT_cauchy_cum}).}
\end{figure}
We repeat the analysis of the Jarzynski estimator as done for the Gaussian example in the previous section and determine the log-PPV $\ln\ppv$ and error margins for the bimodal likelihood distribution defined in (\ref{eq:FAT_cauchy_bimodal}). To do so, we generate $N=10^8$ Markov chains using again the MCMC algorithm and compute the corresponding $R$-values from (\ref{eq:defR}).
First, we demonstrate in Fig.\ \ref{fig:FAT_var_expR_conv} that the variance of the random variable $\exp(R)$ is finite, as required for the central limit theorem to be applicable. Next, Fig.\ \ref{fig:FAT_gore} reveals that the central limit theorem holds for a number of more than about $10^6$ Markov chains, cf. the discussion of the Gaussian example after Fig.\ \ref{fig:cmn} in the previous subsection.

\begin{figure}
\includegraphics[width=0.95\columnwidth]{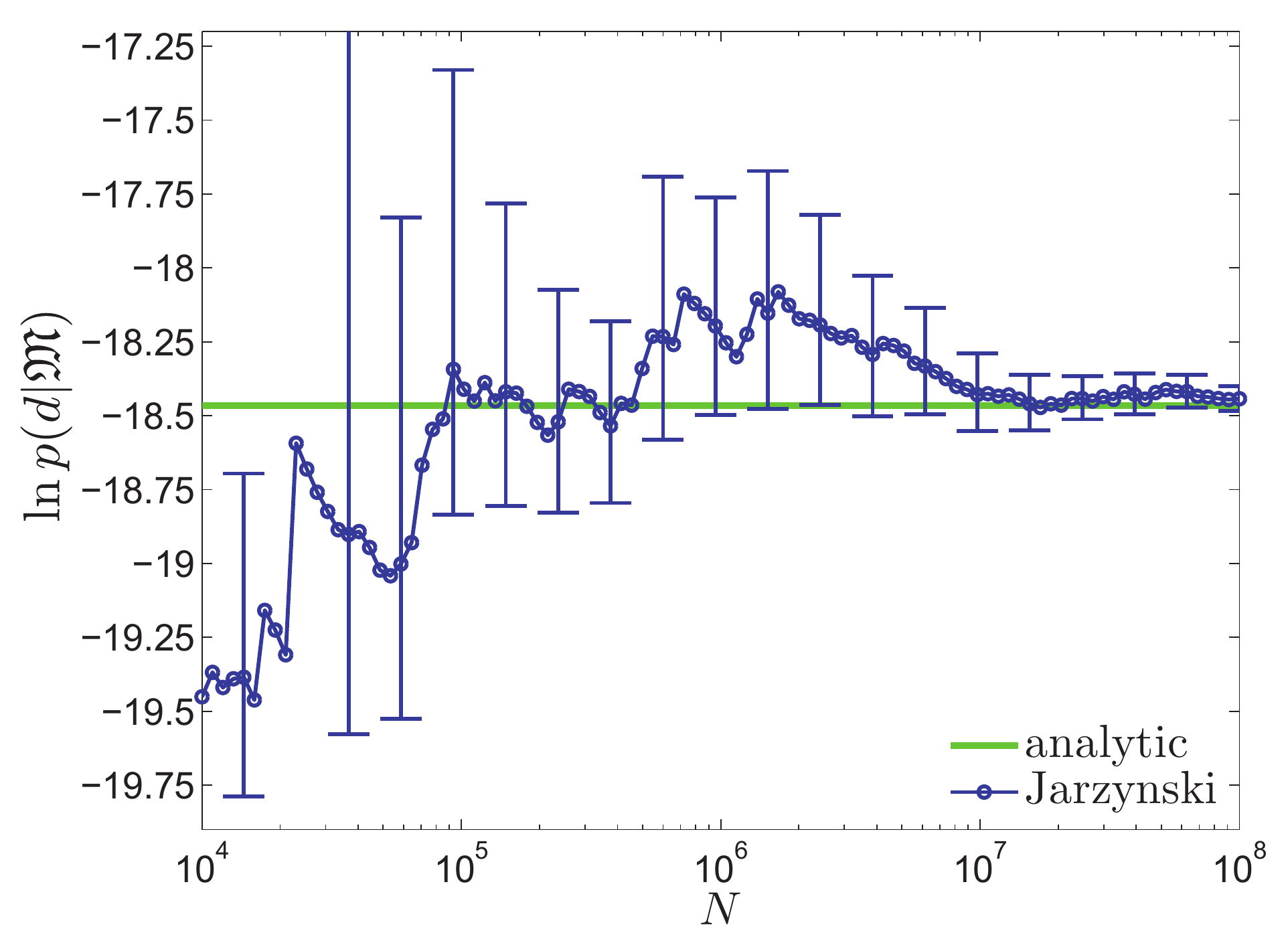}%
\caption{\label{fig:FAT_lnPPV_conv_Nrow1} Estimation of the log-PPV for the likelihood distribution (\ref{eq:FAT_cauchy_bimodal}) involving two Cauchy distributions using the Jarzynski estimator (\ref{eq:Jest_ln}) for an increasing number $N$ of $R$-values. The thick line is the analytic result obtained from (\ref{eq:FAT_cauchy_cum}), the symbols use the Jarzynski estimator. The error bars are given by the confidence limits $\hat D_{\pm}(N)$ from (\ref{eq:dpm_hat}).}
\end{figure}
The confidence limits of the bias of the best estimate for $\ln\ppv$, being $\hat D_\pm(N)$ from (\ref{eq:dpm_hat}), is depicted in Fig.\ \ref{fig:FAT_bias} for an increasing number $N$ of $R$-values, together with $D(N)$ from (\ref{eq:d}) using the exact result of $\ppv$ from (\ref{eq:FAT_cauchy_cum}). It is evident, that for $N>10^6$ the confidence limits $\hat D_\pm(N)$ smoothly approach zero enclosing $D(N)$.

Finally, in Fig.\ \ref{fig:FAT_lnPPV_conv_Nrow1}, we demonstrate the performance of the Jarzynski method and the proposed error analysis for increasing $N$. It is evident that $\hat D_{\pm}$ is again a well suited error margin even for this example of a heavy tailed likelihood distribution, as the true value $\ln\ppv$ is again always covered by $\hat D_{\pm}$.


\section{Averages with the posterior distribution}\label{sec:post}

We now focus on the problem of computing averages with respect to the posterior distribution numerically. Our aim is to investigate the fast-growth algorithm based on \eqref{eq:postav}, which is closely related to the Jarzynski prior-predictive value estimator. We demonstrate that the fast-growth calculations of $\langle\dots\rangle_{\post}$ are particularly advantageous when $p_{\post}(x\vert d,\mathfrak{M})$ is multimodal. The severe problems that, under these circumstances, affect the standard Monte Carlo method are, to a large extent, overcome by the algorithm based on \eqref{eq:postav}.  

For the assessment, we make use of the bimodal Gaussian example described in Sec.\ \ref{sec:simbim}, and consider the average of the function
\begin{equation}
f(x)=x_{\parallel}=\frac{x\cdot d}{\vert d\vert}\label{eq:testf}
\end{equation}
with respect to the posterior distribution. The scalar $x_{\parallel}$ is the component of the vector $x$ along the vector $d$, specifying the locations of the maxima in the posterior distribution. Our simulations are compared with the analytic result
\begin{equation}
\langle x_{\parallel} \rangle_{\post}=(q_{1}-q_{2})\Bigg(\frac{\sigma_{p}^2}{\sigma_{p}^{2}+\sigma_{l}^{2}}\Bigg)\vert d\vert\,,
\end{equation} 
which, for the parameter values used in Sec.\ \ref{sec:simbim}, gives 
\begin{equation}
\langle x_{\parallel} \rangle_{\post}\approx -20.0308\,.\label{eq:suban}
\end{equation}

\begin{figure}
  \centering
   \includegraphics[width=0.97\columnwidth]{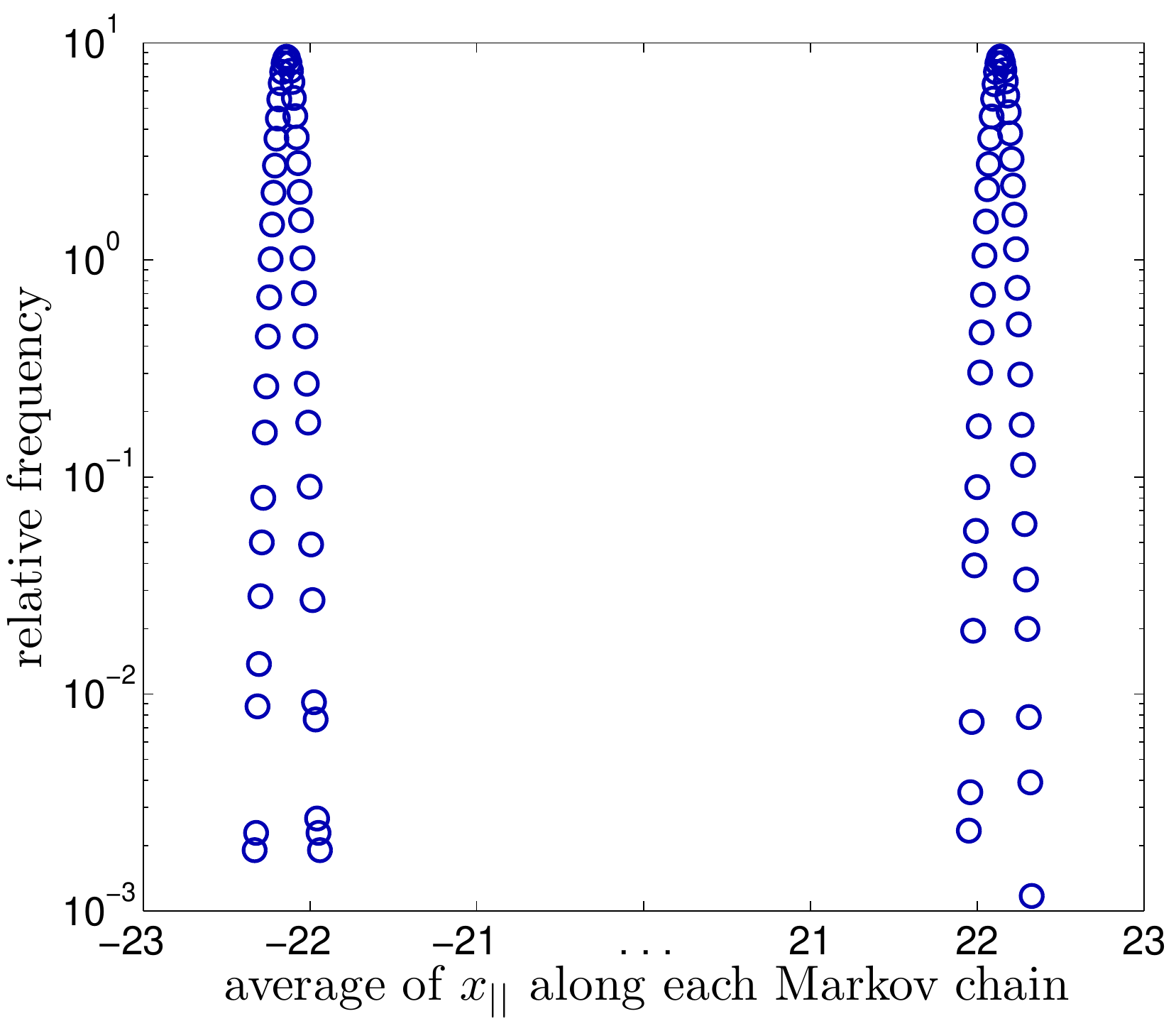}
   \caption{\label{fig:GAU_montecarlohst} Histogram for the average of $x_{\parallel}$ along each Markov chain in our standard Monte Carlo program. Despite their different weights, the peaks of the bimodal posterior distribution are sampled almost equally.}
\end{figure}
To gain insight, it is useful to examine a standard Monte Carlo algorithm. Multiple stationary Markov chains are set to explore the parameter space, with $p_{\post}(x\vert d,\mathfrak{M})$ as their target distribution. Along each trajectory, the average of $x_{\parallel}$ is calculated. Then, a further average across the Markov chains yields an estimate of $\langle x_{\parallel}\rangle_{\post}$. In our simulation, $6\times 10^5$ trajectories are generated, each with $5\times 10^4$ steps. This makes a total of $3\times 10^{10}$ steps, and corresponds to the estimate
\begin{equation}
\langle x_{\parallel} \rangle_{\post}\approx -0.12\,.\label{eq:estmc}
\end{equation}
It is evident that the standard Monte Carlo algorithm fails to solve the problem at hand. The histogram in Fig.\ \ref{fig:GAU_montecarlohst} explains the reason of such failure. Although the two peaks of the posterior distribution have different weights, $q_{1}=\frac{1}{21}$ and $q_{2}=\frac{20}{21}$, they contribute to \eqref{eq:estmc} roughly in equal measure. More specifically, one can determine that the chains get trapped around the maxima of $p_{\post}(x\vert d,\mathfrak{M})$.

Let us investigate the fast-growth estimator
\begin{equation}
\langle x_{\parallel} \rangle_{\post}\approx \frac{\langle e^{R}  x_{\parallel}(T) \rangle_{N}}{\langle e^{R} \rangle_{N}}\,,\label{eq:estfast}
\end{equation}
see \eqref{eq:postav} and \eqref{eq:testf}. As before, $\langle \dots\rangle_{N}$ indicates an empirical average over $N$ non-stationary Markov chains. We consider the same pool of data as in Sec.\ \ref{sec:simbim}. Thus, $N=6\times 10^{7}$, and each trajectory is made up of $500$ steps. Accordingly, both the fast-growth and Monte Carlo simulations include $3\times 10^{10}$ steps in total, which produces similar running times.

\begin{figure}
  \centering
   \includegraphics[width=0.97\columnwidth]{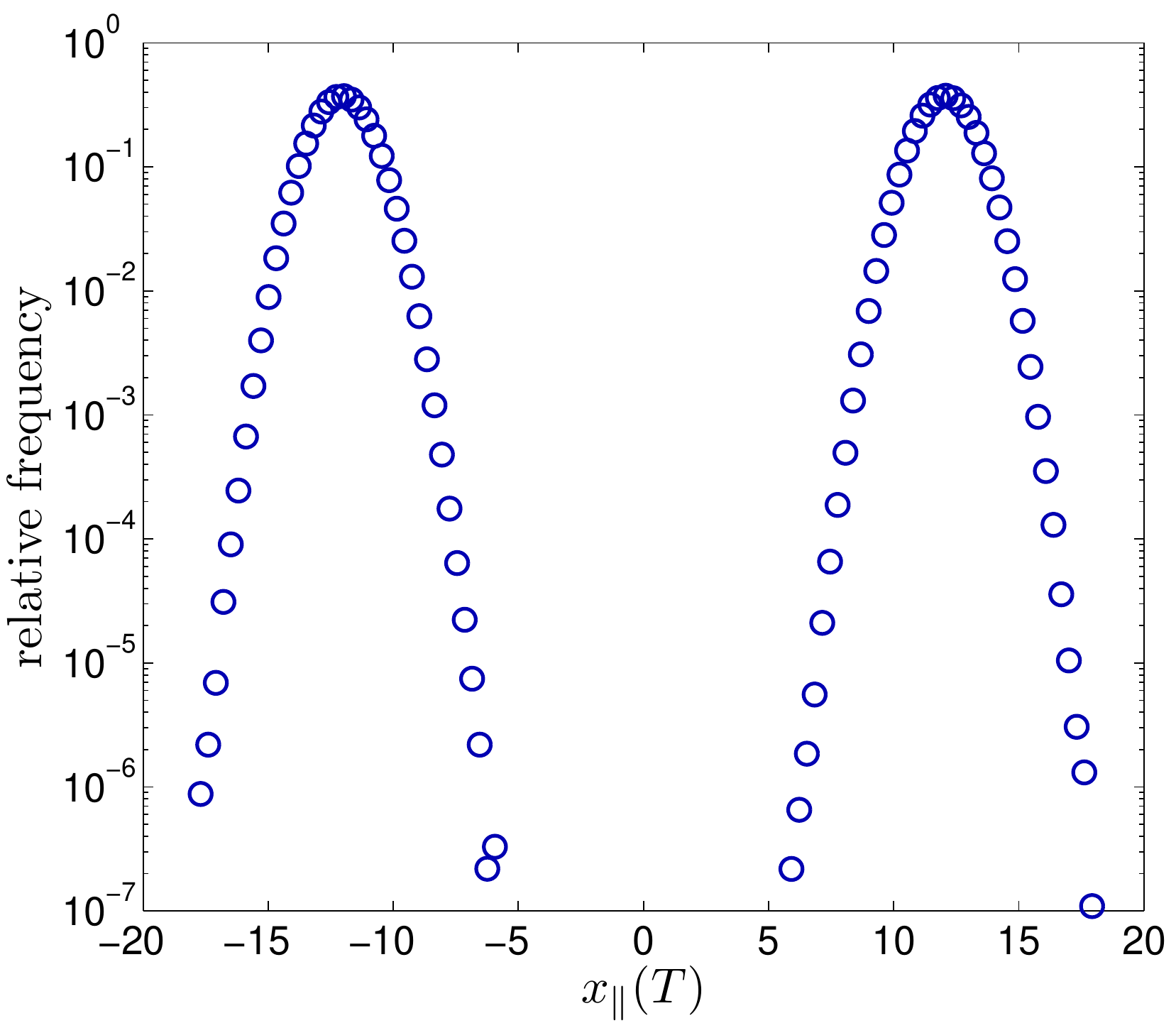}
   \caption{\label{fig:GAU_fasthst} Histogram for the end-location $x_{\parallel}(T)$ of each trajectory in our fast-growth program. Although the Markov chains are non-stationary, they still get trapped around the maxima of the posterior distribution.}
\end{figure}
The absolute error in the fast-growth calculation, 
\begin{equation}
\left\vert\frac{\langle e^{R}  x_{\parallel}(T) \rangle_{N}}{\langle e^{R} \rangle_{N}}-\langle x_{\parallel} \rangle_{\post}\right\vert \approx 1.19 \times 10^{-3}\,,\label{eq:fastbest}
\end{equation}
demonstrates that the method performs well. As a matter of fact, one obtains the estimate $\langle x_{\parallel} \rangle_{\post}\approx -20.0296$. Notably, the histogram in Fig.\ \ref{fig:GAU_fasthst} is qualitatively rather similar to the one in Fig.\ \ref{fig:GAU_montecarlohst}. Even if the Markov chains for the fast-growth algorithm are non-stationary, the mismatched peaks of the posterior distribution are sampled equally. However, weighing the final value of $x_{\parallel}$ with $e^R$ of the respective trajectory in the estimator \eqref{eq:estfast} resolves the different weights of the peaks in the posterior distribution and yields the correct result for the average. 

\begin{figure}
  \centering
   \includegraphics[width=0.97\columnwidth]{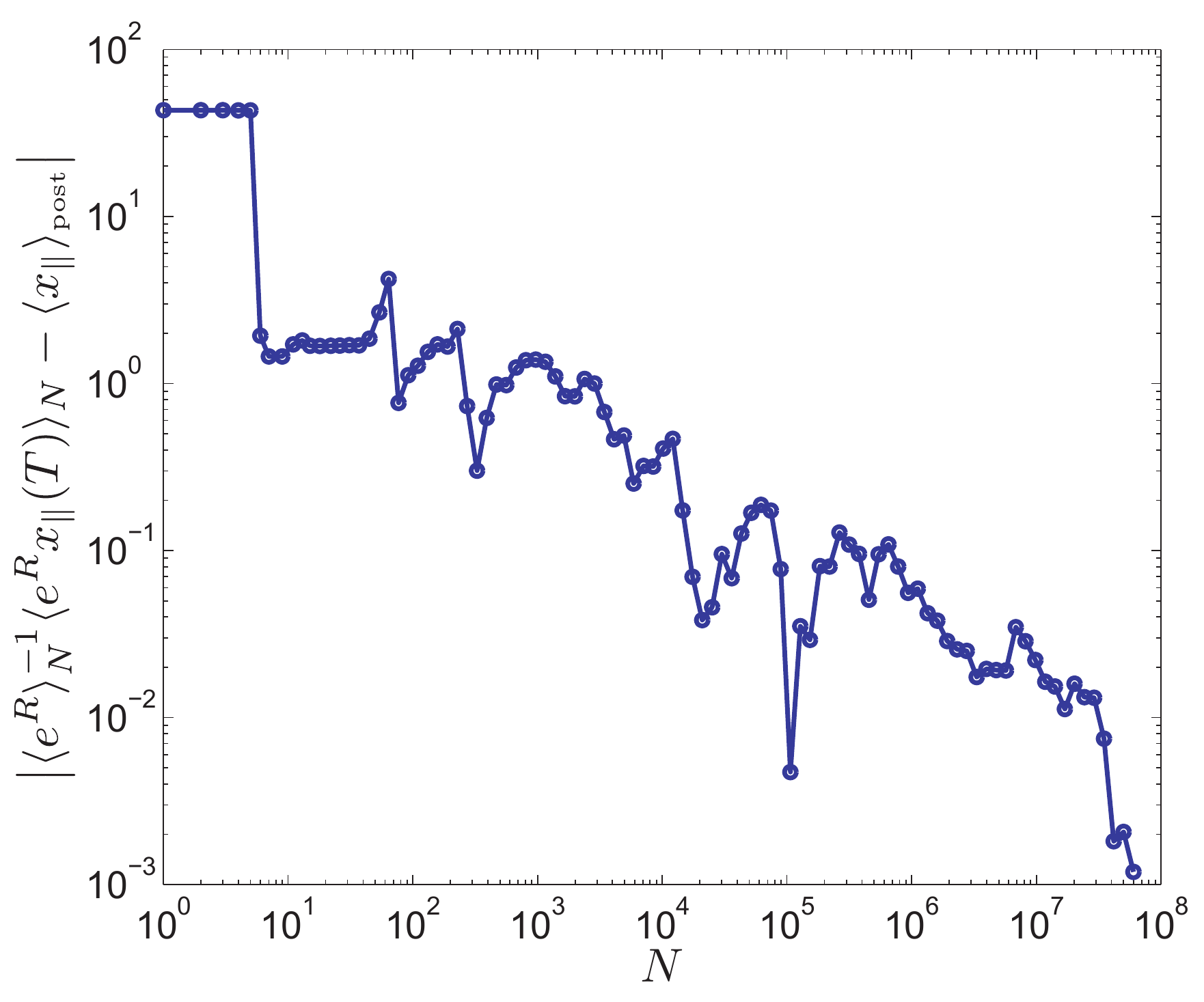}
   \caption{\label{fig:GAU_fast} Absolute error in the fast-growth estimate of $\langle x_{\parallel}\rangle_{\post}$ as the number of trajectories $N$ is varied. The rightmost datapoint corresponds to \eqref{eq:fastbest}.}
\end{figure}
Fig.\ \ref{fig:GAU_fast} specifies the convergence of the fast-growth algorithm as $N$ increases. The detailed error analysis is a topic for future work.


\section{Summary}\label{sec:conc}
Successful use of Bayesian methods in realistic problems of statistical data analysis requires efficient ways to numerically calculate high-dimensional integrals. Due to the similarity of this problem with the determination of free-energy differences of complex molecules, the transfer of methods from statistical mechanics to Bayesian statistics has a long tradition. Notably, thermodynamic integration, which replaces the determination of a normalization factor by an integral over much more accessible averages, has proven very valuable in this connection. 

However, relying on well-equilibrated averages for different temperatures, thermodynamic integration runs into difficulties in the presence of multimodal distributions. Since multimodal likelihoods and posterior distributions are quite common in Bayesian data analysis, a method less dependent on perfect equilibration is called for. In statistical mechanics, the Jarzynski equation and the Crooks relation have been used successfully to determine free-energy differences from non-equilibrium trajectories without final relaxation. Slightly modified variants of these relations may be implemented to determine the prior-predictive value and posterior averages respectively in Bayesian statistics. 

In the present paper we have performed a detailed analysis of the statistical error inherent in these methods. From the determination of free-energy differences with the help of the Jarzynski equation it is know that the method has a bias due to the involved non-linearities. To keep track of this bias in the setting of Bayesian data analysis, we have split the mean-square error of the estimator into a contribution from the bias and from the variance. As usual, the variance may be well characterized by the empirical sample variance, whereas the bias depends on the exact value of the prior-predictive value which is not known. We have therefore split the bias once more into a contribution that, similarly to the variance, may be characterized by the sample data alone, and a remainder for which we provide bounds in form of a confidence interval. Taking everything together, we finally give a confidence interval for the prior-predictive value determined from instationary Markov chain Monte-Carlo simulations which for multimodal distributions are superior to thermodynamic integration. 

We have tested our results against extensive numerical simulations of two model cases with bimodal likelihoods. These are either sums of two Gaussians or of two Lorentzians. Combined with appropriate prior distributions, the prior-predictive values can be calculated analytically for both cases which facilitates the comparison with the simulation results. By investigating various samples sizes $N$, our analytical findings were all verified, and the  predicted dependence of the error measures on $N$ was reproduced. Our results are also consistent with error measures discussed previously in connection with free-energy estimates. Similarly, agreement was found for the determination of averages with multimodal posterior distributions using the Crooks relation, where straight Monte-Carlo sampling of the posterior was seen to be problematic.

In conclusion, variants of the recently discovered fluctuation theorems of non-equilibrium statistical mechanics may prove very helpful in Bayesian data analysis if multimodal distributions are relevant. In these cases, they allow an efficient determination of high-dimensional integrals via Markov chain Monte-Carlo methods without requiring complete equilibration. Admittedly, these methods build on exponential averages which may converge poorly and which show a bias that needs to be monitored. As in statistical mechanics, the  trade-off between problems of equilibration and subtleties of exponential averages is difficult to assess in general and has to be analyzed for each case at hand individually.

\section*{Acknowledgement}
Financial support from the DFG under project EN 278/6-1 is gratefully
acknowledged.

\bibliography{ppv.bib}

\end{document}